\documentclass[aps,prb,twocolumn,superscriptaddress]{revtex4-2}
\usepackage{graphicx,bm,amsmath,amssymb,url,hyperref, braket, color}

\begin{document}
\title{Measuring the arrival time of an electron wave packet using a dynamical potential barrier}

\author{Wanki Park}
\affiliation{Department of Physics, Korea Advanced Institute of Science and Technology, Daejeon 34141, Korea}

\author{H.-S. Sim} 
\affiliation{Department of Physics, Korea Advanced Institute of Science and Technology, Daejeon 34141, Korea}

\author{Sungguen Ryu} \email[]{sungguen@ifisc.uib-csic.es}
\affiliation{Department of Physics, Korea Advanced Institute of Science and Technology, Daejeon 34141, Korea}
\affiliation{Institute for Cross-Disciplinary Physics and Complex Systems IFISC (UIB-CSIC), E-07122 Palma de Mallorca, Spain}

\date{\today}

\begin{abstract}

A time-dependent potential barrier has been used to probe the arrival-time distribution of the wave packet of a hot electron by raising the barrier to block the packet upon arrival of the packet at the barrier. To see whether the barrier precisely detects the distribution, it is necessary to study an error caused by a finite rising speed of the barrier. For this purpose, we study transmission of an electron wave packet through the dynamical barrier, and identify two regimes, the semiclassical regime and the quasistatic regime. In each regime, we calculate the arrival-time distribution reconstructed by using the barrier and quantify the error in the detection, the difference of the temporal uncertainty between the wave-packet distribution and the reconstructed distribution. Our finding suggests that for precise detection, the time scale, in which the barrier height rises over the energy distribution of the wave packet and the tunneling energy window of the barrier, has to be much shorter than the temporal uncertainty of the wave packet. The analytical results are confirmed with numerical calculations.
\end{abstract}

\maketitle

\section{Introduction}

Measuring a single-electron state in nanoelectronic devices is an important task in electron quantum optics~\cite{bocquillon2012electron,bocquillon2013coherence,bauerle2018coherent} and quantum technology. For low-energy single-electron sources such as the mesoscopic capacitor~\cite{feve2007demand} and the Leviton~\cite{dubois2013minimal}, a quantum homodyne tomography~\cite{jullien2014quantum,bisognin2019quantum} has been demonstrated.
The tomography is based on Hong-Ou-Mandel-type collision of the low-energy excitations on the Fermi sea.

On the other hand, single-electron sources of hot electrons such as a quantum-dot pump~\cite{blumenthal2007gigahertz, giblin2012towards} generate hot electrons of energy of typically $\sim$100 meV above the Fermi sea~\cite{kataoka2017time}.
Unlike the electrons near the Fermi energy, the hot electrons are effectively isolated from the other electrons in the Fermi sea and phonons, with the mean-free path larger than micrometers in the strong magnetic field condition~\cite{taubert2011,emary2016,johnson2018phonon, ota2019}.
For such hot electrons, a continuous-variable tomography~\cite{fletcher2019continuous,locane2019time} was recently suggested. This is based on time-energy filtering by a tunable potential barrier.
Depending on whether the barrier height remains static or rises, the potential barrier provides either a detector for energy spectroscopy or an arrival-time detector for the hot electrons~\cite{fletcher2013clock, kataoka2016time, kataoka2017time}.

The time resolution of the arrival-time detection by the dynamical potential barrier is nontrivial. For example, the electron can be excited through a photoassisted process~\cite{ikeda} while transmitting through the barrier. Moreover, there is the traversal time~\cite{buttiker1982traversal} during the transmission. It is hence desirable to have a quantitative theoretical study in which the shape, rising speed, and tunneling time of a realistic potential barrier are taken into account. We note that the rising speed is often limited in experimental situations~\cite{kataoka2017time}, as the dynamical potential barrier can induce side effects diminishing the accuracy of a single-electron source nearby via heating effects or crosstalks through other electronic gates~\cite{ryu2021rf}.

\begin{figure}[t]
\includegraphics[width=.95\columnwidth]{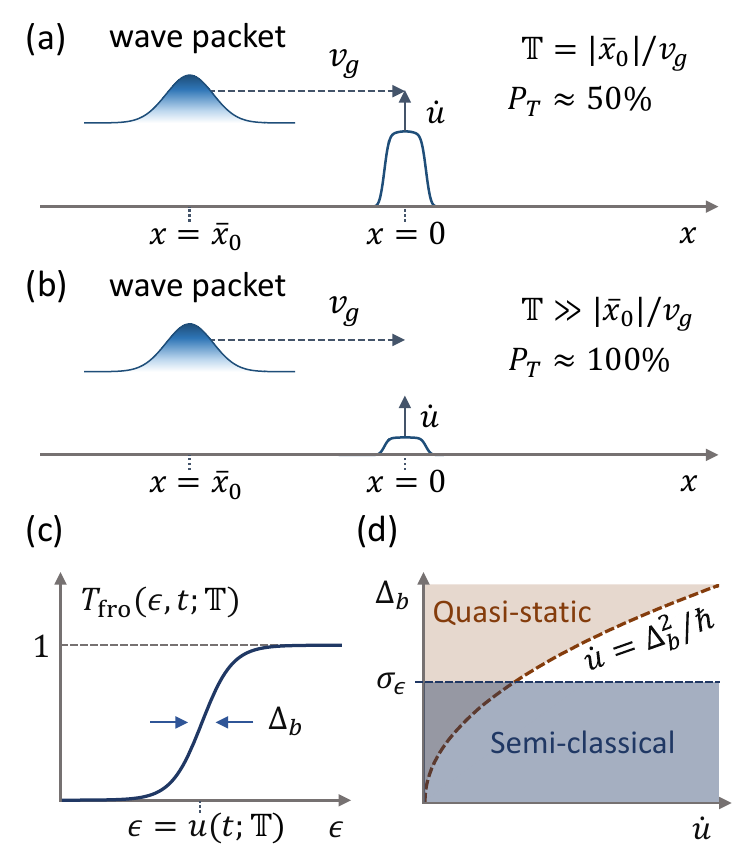}
\caption{(a), (b) Protocol for measuring the arrival-time distribution $P_{\text{ATD}}$ of a single-electron wave packet. A wave packet propagates towards a potential barrier (the ATD detector) with group velocity $v_g$. The barrier height increases with rate $\dot{u}$. Depending on the height at the moment of the arrival, the wave packet transmits through the barrier with transmission probability $P_T \approx 50 \%$ [case (a)], or with $P_T \approx 100 \%$ [case (b)].
The height at the arrival is tuned by applying a delay $\mathbb{T}$ to the time-dependent voltage operating the barrier.
The derivative $\partial P_T/\partial \mathbb{T}$, which is measurable in experiments, reconstructs $P_{\text{ATD}}$ under certain conditions.
(c) Transmission probability $T_\textrm{fro}$ of a plane wave of energy $\epsilon$ through a static barrier with the height $u(t;\mathbb{T})$.
(d) Quasistatic and semiclassical regimes.}
\label{fig:setup}
\end{figure}

The goal of this study is to clarify the resolution of the arrival-time detection. We introduce the measurement protocol~\cite{fletcher2013clock}; see Figs.~\ref{fig:setup}(a) and \ref{fig:setup}(b). The height of the potential barrier, which detects the arrival time at the position of the barrier, quickly rises in time while the wave packet propagates towards the barrier. 
The transmission probability $P_T$ of the packet through the barrier depends on the barrier height at the moment of the arrival.
The height at the arrival is tuned by applying a delay $\mathbb{T}$
which shifts the time-dependent voltage $V(t)$ operating the barrier to $V(t-\mathbb{T})$.
The wave-packet transmission probability $P_T$ is read out for different values of the delay in experiments by measuring electric currents through the barrier.
Then, one approximately reconstructs the arrival-time distribution by calculating the derivative of the transmission probability  with respect to the delay, $\tilde{P}_\text{ATD}\equiv \partial P_T/\partial \mathbb{T}$. In the limit where the rising speed of the barrier goes to infinity, the reconstructed arrival-time distribution (R-ATD) $\tilde{P}_\text{ATD}$ becomes identical to the arrival-time distribution (ATD) $P_{\text{ATD}}$~\cite{ryu2016ultrafast}. 

In this work, we analytically and numerically study the dynamics of a single-electron wave packet injected to a dynamical potential barrier. We quantify the error in detecting the ATD that originates from a finite rising speed of the barrier height, comparing the R-ATD at a finite rising speed and the ATD. 
We find the condition for successful reconstruction of the ATD that the timescale $\text{max}(\Delta_b, \sigma_\epsilon)/\dot{u}$ for the barrier to block the wave packet energetically should be much shorter than the temporal uncertainty of the wave packet,
where $\Delta_b$ is the energy scale over which the transmission probability of a plane wave through the barrier changes from 0 to 1, $\sigma_\epsilon$ is the wave-packet energy uncertainty, and $\dot{u}$ is the rate of rise of the barrier height. Furthermore, we identify two regimes of the transmission of the wave packet through the dynamical potential barrier, the semiclassical regime and the quasistatic regime, and find a simple relation between the time uncertainties of the ATD and the R-ATD, as a quantitative indicator of the error, in each regime.

This paper is organized as follows. In Sec.~\ref{sec:Setup}, we introduce a model for the dynamical potential barrier, the arrival time distribution, and the two regimes. 
In Sec.~\ref{sec:SC},  we study transmission of a wave packet through the dynamical potential barrier in the semiclassical regime. Applying an approximation suitable for the regime, we obtain an analytic relation between the ATD and the R-ATD. 
In Sec.~\ref{sec:QS}, we obtain the corresponding relation in the quasistatic regime. The conclusion and discussion are given in Sec.~\ref{sec:Conclusion}.

\section{Model}\label{sec:Setup}

In this section, we explain the model for the dynamical potential barrier, the arrival-time distribution, and the two regimes of the transmission of a hot-electron wave packet through the barrier.

\subsection{Dynamical potential barrier}
\label{sec:potential}

In the ATD measurement of Refs.~\cite{fletcher2013clock, waldie2015measurement, kataoka2017time}, a hot-electron wave packet is generated by a quantum-dot pump formed in a two-dimensional electron system under a strong perpendicular magnetic field. It travels along a quantum Hall edge, and it is detected at an ATD detector which is a time-dependent potential barrier formed on the edge.
The scattering of the emitted electron at the barrier is well described by the saddle-point constriction model due to the strong magnetic field~\cite{fertig1987transmission,buttiker1990quantized}.
According to the model, the transmission probability of the emitted electron is determined by the tunneling through an inverse harmonic potential barrier.
Hence, we consider a one-dimensional Hamiltonian  [see Fig.~\ref{fig:setup}(a)]
(see Appendix \ref{sec:eff_1D} for its relation to the two-dimensional electron system)
\begin{equation}
H (x,t) = - \dfrac{\hbar^2}{2m^*} \dfrac{\partial^2}{\partial x^2} + U(x,t; \mathbb{T}).
\label{eq:Hamiltonian}
\end{equation}
Here $m^*$ is the effective electron mass and
$U(x,t;\mathbb{T})$ is an inverse harmonic potential with a barrier smoothing away from the top,
\begin{equation}
U(x, t;\mathbb{T}) = f_{s}(x) \bigg[ u(t;\mathbb{T}) -\frac{1}{2}m^*\omega_b^2 x^2 \bigg] .
\label{eq:potential_inverse_harmonic}
\end{equation}
$u(t;\mathbb{T})\equiv (t-\mathbb{T}) \dot{u} +\bar{\epsilon}$ is the barrier height at the barrier center position $x=0$. The barrier height linearly increases over time $t$ with a rate $\dot{u}$, $\mathbb{T}$ is the time delay applied to the barrier~[Figs.~\ref{fig:setup}(a) and \ref{fig:setup}(b)]. $\bar{\epsilon}$ is the mean energy of the wave packet.
The term of $-(1/2)m^*\omega_b^2 x^2$ dictates that the barrier near the top is in the form of the inverse harmonic potential of energy scale $\hbar\omega_b$, or equivalently length scale $l_b = \sqrt{\hbar/(m^* \omega_b)}$. 
The parameter $\omega_b$ determines the transmission probability $T_\text{fro}$ 
\begin{equation}
T_\text{fro}(\epsilon, t;\mathbb{T})= \frac{1}{1+ \exp[-\pi\{\epsilon -u(t;\mathbb{T})\}/(\sqrt{3}\Delta_b)]}
\end{equation}
of a plane wave of energy $\epsilon$ 
through a static barrier whose potential shape is identical to that of the dynamical barrier frozen at time $t$,
where $\Delta_b =\hbar\omega_b/(2\sqrt{3}) $ [Fig.~\ref{fig:setup}(c)].
$f_{s}(x) $ is the smoothing factor which makes the potential $U$ vanish in the region of $|x| \ge l_{s}$ far away from the barrier center with the smoothing length $l_{s}$.
  As long as the factor $f_s$ does not alter the inverse harmonic behavior around the barrier top within the length scale $l_b$, the detailed choices about $f_s$ do not affect our result about R-ATD because the packet transmission probability is only determined by the potential near the barrier top~\cite{fertig1987transmission}.
  To achieve this condition, $l_s$ should be much larger than $l_b$, and the Taylor expansion of $f_s$ at $x=0$ should be $1 + O(x/l_s)^4$.
For the numerical results below, we use $f_s(x) =1/\text{cosh}[ (2x/l_s)^4]$ and $l_s \sim 10 l_b$. Note that the smoothing factor $f_s$ satisfies the above condition. The choice for $l_s$ is based on the experimental situations. The distance between the two-dimensional electron system and the top gates is $\sim$100 nm, which tends to smooth out~\cite{davies1995modeling} the electrostatic potential induced by the gates by the same length scale. The length scale of the inverse harmonic confinement $l_b$ is in the order of the magnetic length (typically $\sim$10 nm for magnetic field of $10$ T).

\subsection{Arrival time distribution}

The single-electron wave packet is emitted from the pump to the one-dimensional channel described by Eq.~(\ref{eq:Hamiltonian}).
We describe this state as $\ket{\psi_0}$ which has mean energy $\bar{\epsilon}$ and position $\bar{x}_0$ (located far away from the barrier) at time $t=0$.
Then, the wave packet propagates toward the barrier with group velocity $v_g=\sqrt{2\bar{\epsilon}/m^*}$~\cite{kataoka2016time}. We apply the condition that the mean energy is much larger than the energy uncertainty of the wave packet since the mean energy and the energy uncertainty are typically $\sim$100 meV and $\sim$1 meV, respectively, in experiments~\cite{fletcher2013clock}.
This condition allows us to use the linear dispersion relation $\epsilon=\bar{\epsilon}+ v_g (p-\bar{p})$ between the kinetic energy $\epsilon$ and momentum $p$ around the mean momentum $\bar{p}$ in the analytic calculations below.
We focus on the case that the emitted electron is in the pure state of the Gaussian wave packet of energy uncertainty (i.e., the standard deviation) $\sigma_\epsilon$ and temporal uncertainty $\sigma_t$.
The Gaussian form is expected when the quantum-dot pump is operated in the strong magnetic field and with the fast emission protocol~\cite{ryu2016ultrafast}.
In Appendix~\ref{sec:Correlated_state}, we discuss the case where the initial wave packet is not Gaussian~\cite{kataoka2017time, fletcher2019continuous} and show that the conclusion about the condition for successful ATD reconstruction remains the same.

It is convenient to describe the initial state by the Wigner distribution.
It can be represented in position $x$ and momentum $p$ as
\begin{equation}
  \label{eq:W-xp}
  W_0(x,p)
  = \dfrac{1}{\pi \hbar} \int^{\infty}_{-\infty} dx' \, \psi_0^*(x+x') \psi_0(x-x') e^{2ip x'/\hbar}, 
\end{equation}
where $\psi_0(x)$ is the wave function of the initial state at position $x$.
Using the linear dispersion relation,
the Wigner distribution can be also represented in energy $\epsilon$ and time of arrival $t_a$ at $x=0$,
\begin{equation}
  \label{eq:W-tE}
\mathcal{W}_0(t_a,\epsilon) \equiv \dfrac{1}{\pi \hbar} \int^{\infty}_{-\infty} d \epsilon' \phi_0^*(\epsilon+\epsilon') \phi_0(\epsilon-\epsilon')e^{2i\epsilon' t_a/\hbar},
\end{equation}
where $\phi_0(\epsilon)$ is the amplitude of the initial state at kinetic energy $\epsilon$.
The two representations are related as $\mathcal{W}_0(t_a, \epsilon) = W_0(-v_gt_a, \epsilon/v_g) $. For Gaussian wave packet, $\mathcal{W}_0$ becomes
\begin{equation}
\mathcal{W}_0(t_a ,\epsilon) = \dfrac{1}{\pi \hbar} \exp \bigg[ - \dfrac{(t_a - \bar{t}_a)^2}{2 \sigma_t^2} - \dfrac{(\epsilon - \bar{\epsilon})^2}{2 \sigma_\epsilon^2} \bigg],
\label{eq:W-Gaussian}
\end{equation}
where $\sigma_\epsilon = \hbar / (2 \sigma_t)$ and $\bar{t}_a = -\bar{x}_0/v_g$.

The ATD at the detector ($x=0$) is determined by the marginal distribution of Wigner distribution,
\begin{equation}
  P_{\text{ATD}}(t) = \int^{\infty}_{-\infty} d\epsilon \, \mathcal{W}_0(t, \epsilon) .
  \label{P_ATD}
\end{equation} 
Note that Eqs.~(\ref{eq:W-tE}) and (\ref{P_ATD}) give $ P_{\text{ATD}}(t)  = v_g |\psi_0(-v_g t)|^2$, namely, the ATD is determined by the spatial distribution of the initial packet.

On the other hand, the R-ATD $\tilde{P}_{\text{ATD}}$, obtained using the time-dependent detector barrier with a delay $\mathbb{T}$, is obtained as
\begin{align}
  P_T (\mathbb{T})&= \lim_{t' \rightarrow \infty} \int_0^\infty dx \, |\psi(x,t';\mathbb{T})|^2 , \label{eq:P_T} \\
\tilde{P}_{\text{ATD}}(t) &= \frac{\partial P_T}{\partial \mathbb{T} }\Big|_{\mathbb{T}\rightarrow t},  \label{eq:P_tilde_ATD}
\end{align}
where $\psi(x,t;\mathbb{T})$ is the transmitted wave packet at position $x>0$ and time $t>0$ through the detector with the delay $\mathbb{T}$ and 
$P_T(\mathbb{T})$ is the transmission probability of such packet.

  Note that the R-ATD is the result of the indirect measurement of the ATD through the protocol of Fig.~\ref{fig:setup}. In other words, R-ATD is analogous to the reconstructed density matrix in quantum tomography which indirectly measures the density matrix by a chosen protocol. Below we compare R-ATD and ATD quantitatively to find the condition for a successful ATD measurement protocol in which R-ATD approaches ATD.

\subsection{Regimes}\label{sec:Regimes}

To analyze $\tilde{P}_{\text{ATD}}$, we identify two regimes, the semiclassical regime and the quasistatic regime. For this purpose, we first explain two timescales characterizing the detector barrier, $\hbar/\Delta_b$ and $\Delta_b/\dot{u}$.
The timescale $\hbar/\Delta_b$ characterizes the traversal time of the barrier~\cite{landauer1994barrier,moskalets2002floquet}.
It is the largest traversal time of the static situation of the barrier,
namely the maximum of
\begin{equation}
  \label{eq:traversal_time}
  \tau_{\text{tra}}(\epsilon, t;\mathbb{T})
  = \hbar \left| \frac{\partial}{\partial \epsilon} \ln{ d_\text{fro}(\epsilon,t; \mathbb{T})} \right|
\end{equation}
over different values of $\epsilon$ and $t$, where $d_\text{fro}(\epsilon, t; \mathbb{T})$ is the transmission amplitude of a plane wave of energy $\epsilon$ through the barrier frozen at $t$.
The largest traversal time $\sim$$\hbar /\Delta_b$ is found when the energy $\epsilon$ is almost aligned with the barrier height. See Appendix~\ref{sec:Traversal_time} for the details.
The time scale $\Delta_b/\dot{u}$ characterizes the time during which the transmission probability $T_\text{fro} = |d_\text{fro}|^2$ changes significantly over $t$.

We classify the regimes of the ATD-measurement protocol in the parameter space of $\Delta_b$ and $\dot{u}$ which characterizes the detector barrier [see Fig.~\ref{fig:setup}(d)].
Comparison between $\dot{u}$ and $\Delta_b^2/\hbar$ indicates whether the rise of the barrier height induces the photoassisted tunneling or not.
The quasistatic regime is achieved when $\dot{u} \ll \Delta_b^2/\hbar$; the time evolution of an electron during its scattering with the barrier follows the quasistatic evolution~\cite{moskalets2004adiabatic, moskalets2005magnetic} as the barrier does not change significantly during the traversal time,
namely, $\dot{u} \tau_{\text{tra}} \ll \Delta_b$ at any instance during the wave-packet transmission and $\tau_{\text{tra}}$ is upper bounded by $\hbar/\Delta_b$.
On the other hand, it is also useful to compare between the energy uncertainty $\sigma_\epsilon$ of the emitted wave packet and the energy broadening $\Delta_b$ of the barrier. 
When $\Delta_b \ll \sigma_\epsilon$, the time evolution is well described by the semiclassical dynamics since only a small portion $\sim O(\Delta_b/\sigma_\epsilon) $ of the wave packet undergoes probabilistic quantum tunneling through the barrier.
We focus on the quasistatic and the semiclassical regimes and discuss the other case in the Conclusion.

\section{Semiclassical regime}\label{sec:SC}

Here we compare the R-ATD to the ATD in the semiclassical regime, i.e.,  $\Delta_b \ll \sigma_{\epsilon}$. We use the semiclassical approximation~\cite{polkovnikov2010phase} for the packet transmission probability (see Appendix~\ref{sec:QC} for the validity of the approximation), which is determined by the initial Wigner distribution and the classical trajectories as
\begin{equation}
  P_T(\mathbb{T}) \simeq \int^{\infty}_{-\infty} dx_0 \int^{\infty}_{-\infty} d p_0 \, W_{0}(x_0,p_0)\, \Theta\big(\lim_{t \rightarrow \infty}x_\text{cl}(t)\big) .
\label{eq:P_T_SC_0}
\end{equation}
$x_\text{cl}(t)$ is the position at time $t$ classically evolved from the initial position $x_0$ and momentum $p_0$. The term $\Theta [\lim_{t\rightarrow \infty} x_{\text{cl}}(t)]$, where $\Theta(\cdots)$ denotes the Heaviside step function, describes whether a classically evolved particle ultimately passes over the barrier or not; it is 1 when the particle passes and 0 otherwise. 

To specify the behavior of the term $\Theta[\lim_{t \rightarrow \infty} x_{\text{cl}}(t)]$, we explain the condition 
\begin{equation}
\label{eq:cond-tp-0}
\lim_{t \rightarrow \infty}x_{\text{cl}}(t) =0.
\end{equation}
In a classical trajectory satisfying the condition, the kinetic energy of the electron vanishes when it arrives at the center of the barrier; otherwise, the electron ends up being away the barrier over further time evolution.
Hence, the initial kinetic energy equals the total work done by the electron, Eq.~(\ref{eq:cond-tp-0}) is equivalent to 
\begin{align}
  \label{eq:cond-tp-1}
  \frac{p_0^2}{2m^*}
  & = \int^0_{x_0} dx \,  \dfrac{\partial}{\partial x} U(x,t; \mathbb{T}) \Big|_{t=t_{\text{cl}}(x)} \\
   & = \int^0_{x_0} dx \, \dfrac{df_{s} (x)}{dx} u(t_{\text{cl}}(x); \mathbb{T}),
  \label{eq:cond-tp-2}
\end{align}
where, $t_{\text{cl}}(x)$, defined by the inverse function of $x_{\text{cl}}(t)$,
is the arrival time at position $x$ for a given initial position $x_0$ and momentum $p_0$, that satisfy Eq.~\eqref{eq:cond-tp-0}. In the second line, we use Eq.~(\ref{eq:potential_inverse_harmonic}), an integration by part $\smallint_{x_0}^0 dx \, x^2 \,d f_{s}/d x = -\smallint_{x_0}^0 dx\, 2x f_{s}(x) $, and $f_{s}(x_0)=0$. We simplify Eq.~\eqref{eq:cond-tp-2} using that (i) there is no potential inside $x \in [x_0, -l_{s}]$, hence, $t_{\text{cl}}(-l_s) = (-l_s-x_0)/v_g$, and (ii) the linearity of $u(t;\mathbb{T})$ over time, $u(t;\mathbb{T}) = \dot{u}(t-\mathbb{T})+ \bar{\epsilon}$, 
\begin{equation}
  \label{eq:cond-tp-fin}
  \frac{p_0^2}{2m^*}  = u( -x_0/v_g -C ;\mathbb{T} ),
\end{equation}
where $C = \smallint_{-l_{s}}^0 dx\, [ l_{s}/v_g -t_{\text{cl}}(x) + t_{\text{cl}}(-l_{s}) ] d f_{s}/dx$ is constant independent of $x_0$ and $p_0$ as $t_{\text{cl}}(x) - t_{\text{cl}}(-l_s)$ is independent of $x_0$ and $p_0$. Using Eqs.~(\ref{eq:P_T_SC_0}) and (\ref{eq:cond-tp-fin}), and ignoring $C$ which only induces the R-ATD to be horizontally shifted from the ATD,
we obtain the packet transmission probability $P_T(\mathbb{T})$ in terms of the Wigner distribution
\begin{equation}
  P_T(\mathbb{T}) = \int^{\infty}_{-\infty} dt_a \int^{\infty}_{-\infty} d\epsilon \, \mathcal{W}_{0}(t_a, \epsilon)\, \Theta(\epsilon - u(t_a;\mathbb{T})) .
\label{eq:P_T_SC}
\end{equation}

\begin{figure}[t]
\centering
\includegraphics[width=\columnwidth]{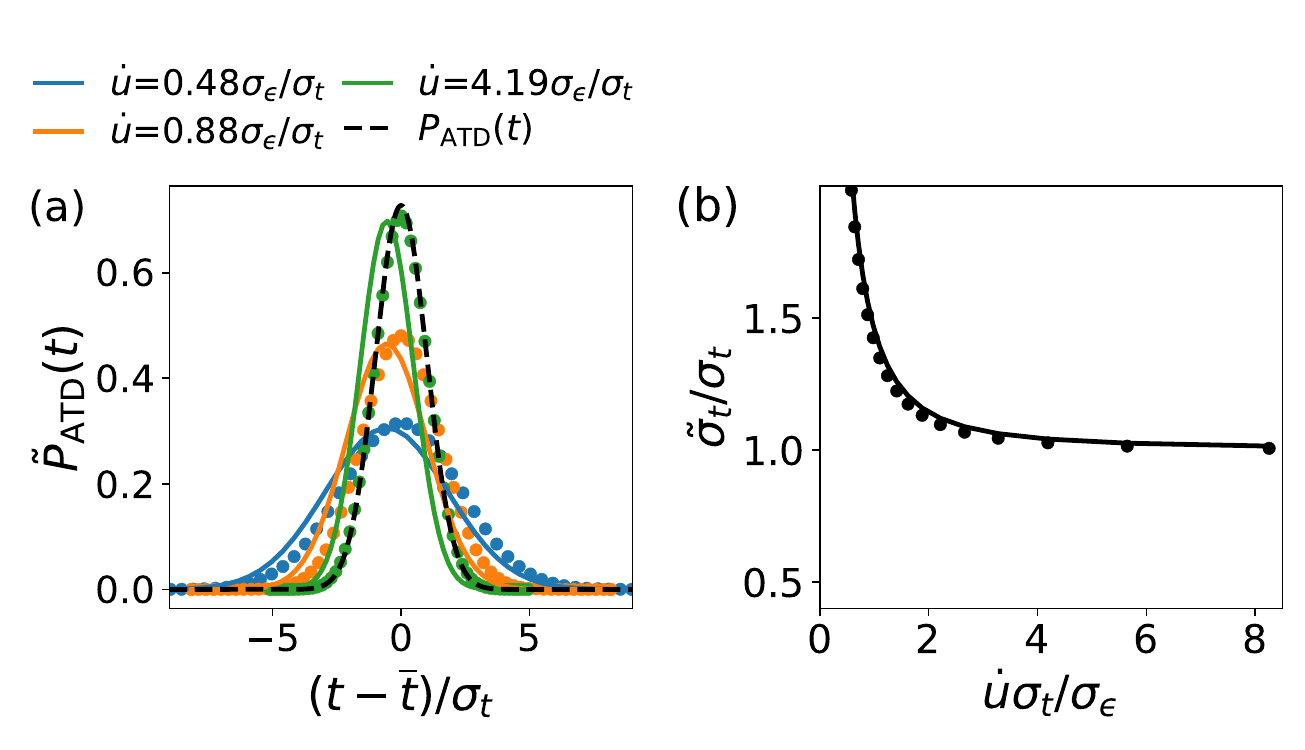} 
\caption{The R-ATD $\tilde{P}_{\text{ATD}}(t)$ in the semiclassical regime.
  (a) $\tilde{P}_{\text{ATD}}(t)$  with various barrier speeds $\dot{u}$ (blue to green). 
  $\tilde{P}_{\text{ATD}}(t)$ approaches $P_{\text{ATD}}(t)$ (dashed) when $\dot{u}> \sigma_\epsilon/\sigma_t$.
  $\bar{t}\equiv -\bar{x}_0/v_g$ is the mean arrival time.
  (b) The standard deviation of $\tilde{P}_{\text{ATD}}(t)$, $\tilde{\sigma}_t$, as a function of $\dot{u}$.
  For both (a) and (b), the solid line denotes the results obtained from the numerical simulation and the circle denotes the semiclassical approximation~\eqref{eq:P_T_SC}.
  Parameters: $\sigma_\epsilon = 0.6 \text{ meV}$ ($\sigma_t = 0.55 \text{ ps}$, respectively) for the energy (time) uncertainty of the initial Gaussian wave packet, and $\Delta_b = 0.1 \text{ meV}$ for the barrier.}
\label{fig:simulation_results_2}
\end{figure}

We finally obtain the R-ATD by using Eqs.~\eqref{eq:P_tilde_ATD} and \eqref{eq:P_T_SC}:
\begin{equation}
  \tilde{P}_{\text{ATD}}(t)
  = \int^{\infty}_{-\infty} d \epsilon  \,  \mathcal{W}_0\Big(t + \frac{\epsilon-\bar{\epsilon}}{\dot{u}},\epsilon \Big) .
\label{eq:P_tilde_ATD_SC}
\end{equation}
$\tilde{P}_{\text{ATD}}$ is determined by the line integral of the Wigner function $\mathcal{W}_0(t_a, \epsilon)$ along the line $t_a = t + (\epsilon - \bar{\epsilon})/\dot{u}$.
  For the ideal detector barrier satisfying $\dot{u}\rightarrow \infty$, the integration gives the marginal distribution in time, hence  $\tilde{P}_{\text{ATD}}(t) \rightarrow P_{\text{ATD}}(t) $.
  For a finite value of $\dot{u}$, a successful reconstruction of the ATD requires that the term $(\epsilon-\bar{\epsilon})/\dot{u}$ should be small in the time and energy window of the Wigner distribution, namely $\sigma_\epsilon/\dot{u} \ll  \sigma_t$.

Plugging the initial Gaussian wave packet of Eq.~\eqref{eq:W-Gaussian} into Eq.~\eqref{eq:P_tilde_ATD_SC}, Eq.~\eqref{eq:P_tilde_ATD_SC} becomes a Gaussian distribution whose standard deviation $\tilde{\sigma}_t$ is broadened compared to that of $P_{\text{ATD}}$ as
\begin{equation}
  \tilde{\sigma}_t = \sqrt{\sigma_t^2 + (\sigma_\epsilon /\dot{u})^2}.
\label{eq:sigma_tilde_SC}
\end{equation}
The equation shows that R-ATD has a larger width than ATD of $\sigma_t$ because of the finite rising time $\sigma_{\epsilon}/\dot{u}$ of the barrier over the packet energy width.

The numerical simulations (see Appendix~\ref{sec:Numerical_simulations} for the method) confirm the analytical analysis with the semiclassical approximation.
Figure~\ref{fig:simulation_results_2}(a) shows the R-ATD $\tilde{P}_{\text{ATD}}(t)$ for various barrier speeds $\dot{u}$ (blue to green).
The results obtained from the numerical simulations (solid lines)
and the semiclassical approximation~(\ref{eq:P_tilde_ATD_SC}) (circle) shows a good agreement. 
$\tilde{P}_{\text{ATD}}(t)$ approaches to $P_{\text{ATD}}(t)$ (dashed line) as the barrier speed becomes larger than $\sigma_\epsilon/\sigma_t$.
Figure~\ref{fig:simulation_results_2}(b) shows that the standard deviation of the R-ATD, $\tilde{\sigma}_t$, approaches to that of the ATD as $\dot{u}$ increases, following the relation (\ref{eq:sigma_tilde_SC}).

\section{Quasistatic regime}\label{sec:QS}

Now we compare the R-ATD to the ATD in the quasistatic regime, i.e., $\dot{u}\ll \Delta_b^2/\hbar$. To describe the wave packet transmitted through the barrier, we first describe the solution $\Psi_{\epsilon}(x,t)$ of the time-dependent Hamiltonian~(\ref{eq:Hamiltonian}), when the incoming state is a plane wave of energy $\epsilon$ coming from the left side of the barrier.
Focusing on the incoming part and transmitted part, and using the quasistatic (also called as adiabatic) approximation~\cite{moskalets2002floquet}, we obtain
\begin{equation}
  \label{eq:psi-sct-QS}
  \Psi_\epsilon (x,t) = 
\begin{cases}
  e^{-\frac{i}{\hbar} \epsilon (t- \frac{x}{v_g})}  + (\text{reflected part}) , & x<0 \\
d_{\text{fro}}(\epsilon, t-x/v_g ;\mathbb{T}) e^{-\frac{i}{\hbar} \epsilon(t- \frac{x}{v_g})} , & x>0.
\end{cases} 
\end{equation}

We expand the initial packet [see Eq.~\eqref{eq:W-Gaussian}] by using the solutions $\Psi_{\epsilon}(x,t)$. Noting that the overlap between the initial packet and $\Psi_{\epsilon}(x,t)$ is determined by the incoming part of $\Psi_{\epsilon}(x,t)$, we obtain the transmitted part of the time-evolved wave packet for $x, t \gg 0$:
\begin{align}
\begin{split}
\psi_T(x,t ;\mathbb{T}) =   & \dfrac{1}{\sqrt{2\pi \hbar v_g}}   \int_{-\infty}^{\infty}  d \epsilon \,   \phi_0(\epsilon) \\ & \times d_{\text{fro}}(\epsilon, t-x/v_g; \mathbb{T}) e^{-\frac{i}{\hbar}\epsilon (t-\frac{x}{v_g})} .
\end{split}
\label{eq:psi_T_QS}
\end{align}
Here $\phi_0(\epsilon)$ is the initial wave function in energy domain $\epsilon$, satisfying the normalization $\smallint^{\infty}_{-\infty} d\epsilon |\phi_0(\epsilon)|^2 = 1$.

The packet transmission probability $P_T$ is obtained when using Eqs.~(\ref{eq:P_T}) and (\ref{eq:psi_T_QS}), and a change of variable from $x$ to $t_a = t-x/v_g $:
\begin{equation}
\begin{aligned}
  P_T (\mathbb{T})
  &=  \frac{1}{\pi \hbar} \int_{-\infty}^{\infty} d t_a \int_{-\infty}^{\infty} d\epsilon \int_{-\infty}^{\infty} d\epsilon'\, 
  \phi_0^*(\epsilon + \epsilon') \phi_0(\epsilon - \epsilon')  \\
  &\qquad \times e^{2 i \epsilon' t_a/\hbar }
    d_\text{fro}^*(\epsilon + \epsilon', t_a;\mathbb{T})
    d_\text{fro}(\epsilon - \epsilon', t_a;\mathbb{T}).
\end{aligned}
\label{eq:P_T_QS}  
\end{equation}
To describe Eq.~\eqref{eq:P_T_QS} in terms of Wigner distribution $\mathcal{W}_0$, we expand transmission amplitudes around the energy $\epsilon$ (with conventions that $0! \equiv 1$ and the zeroth-order derivative of a function is defined as the function itself),
\begin{equation}
  d_{\text{fro}}(\epsilon \pm \epsilon', t_a; \mathbb{T})  
  =    \sum_{n =0 }^{\infty} \frac{(\pm 1)^n}{n!} (\epsilon')^n
  \frac{\partial^n d_{\text{fro}}(\epsilon, t_a;\mathbb{T} )}{\partial \epsilon^n} ,
\end{equation}
and use a property of the Wigner distribution [see Eq.~(\ref{eq:W-tE})]
\begin{equation}
  \big(\frac{\hbar}{2i}\big)^n \frac{\partial^n \mathcal{W}_0(t_a, \epsilon)}{\partial t_a^n}
  =  \int \frac{d\epsilon'}{\pi\hbar} \, (\epsilon')^n
  \phi_0^*(\epsilon + \epsilon') \phi_0(\epsilon - \epsilon')  e^{2 i \epsilon't_a/\hbar }.
\end{equation}
Further using the integration by parts which eliminate the time derivatives applied to the Wigner distribution,
we find that Eq.~\eqref{eq:P_T_QS} becomes
\begin{equation}
  \begin{aligned}
    P_T(\mathbb{T})
      &=\sum_{n,m=0}^{\infty}\frac{(-1)^m}{(2i)^{n+m} n!m!}
      \int_{-\infty}^{\infty} \hspace{-0.3cm} dt_a \int_{-\infty}^{\infty} \hspace{-0.3cm} d\epsilon\, \mathcal{W}_0(t_a, \epsilon) \\ 
    &\quad \times 
     \hbar^{n+m} \frac{\partial^{n+m} }{\partial t_a^{n+m}} \Big[
      \frac{\partial^n d^*_{\text{fro}}(\epsilon,t_a ; \mathbb{T})}{\partial \epsilon^n}
      \frac{\partial^m d_{\text{fro}}(\epsilon,t_a ; \mathbb{T})}{\partial \epsilon^m} \Big].
  \end{aligned}
\end{equation}
The order of each term with $n$, $m$ follows $O[(\hbar \dot{u}/\Delta_b^2)^{n+m} ]$ because the transmission amplitude changes significantly in the timescale of $\Delta_b/\dot{u}$ and energy scale of $\Delta_b$. Hence, in the quasistatic regime of $\hbar \dot{u} / \Delta_b^2 \ll 1$, only the term with $n=m=0$ contributes, and the packet transmission probability follows:
\begin{equation}
  P_T (\mathbb{T})
  = \int_{-\infty}^{\infty} dt_a \int_{-\infty}^{\infty} d\epsilon\, \mathcal{W}_{0}(t_a,\epsilon) T_{\text{fro}}(\epsilon, t_a;\mathbb{T}) .
  \label{eq:P_T_QS_2}
\end{equation}

\begin{figure}[t]
\centering
\includegraphics[width=\columnwidth]{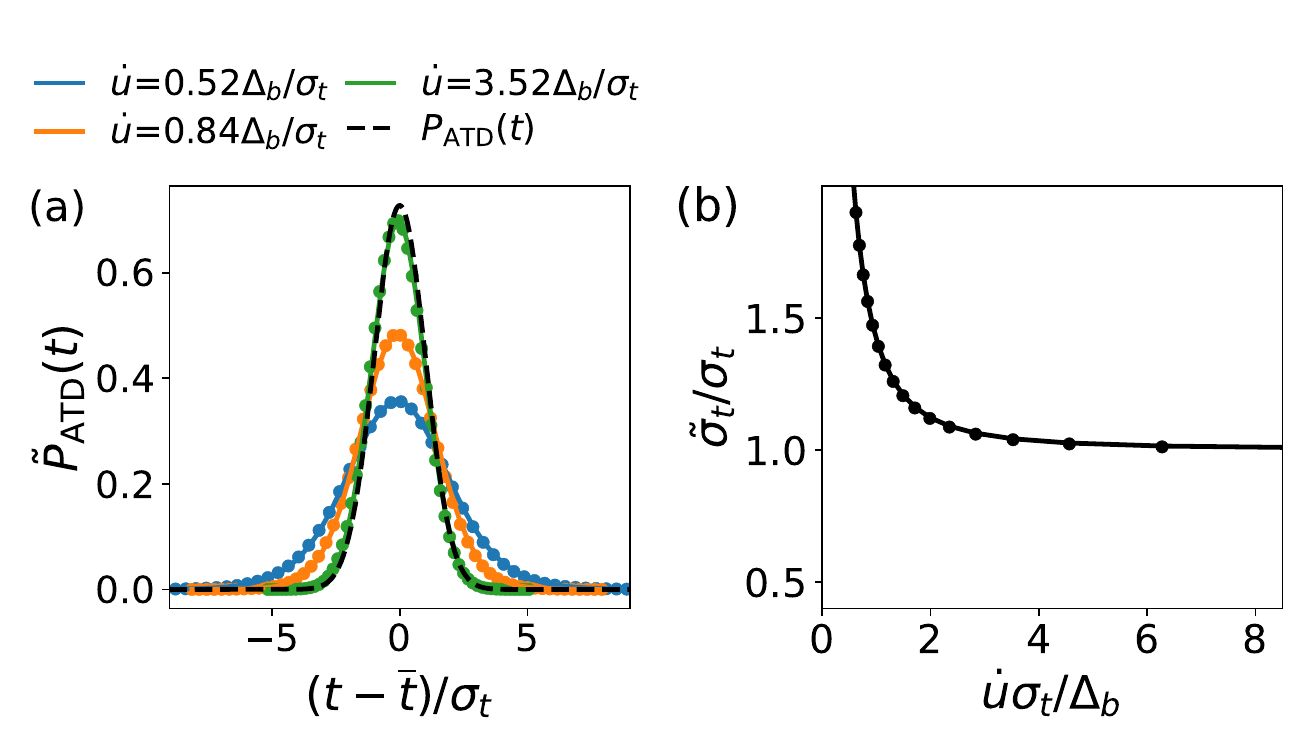} 
\caption{The R-ATD $\tilde{P}_{\text{ATD}}(t)$ in the quasistatic regime.
  The representation and the parameters are the same as Fig.~\ref{fig:simulation_results_2} except that the barrier speed $\dot{u}$ is compared with $\Delta_b/\sigma_t$ where $\Delta_b =6 \text{ meV} \gg \sigma_\epsilon$, and  circles denote the quasistatic approximations (\ref{eq:P_tilde_ATD_QS_2}) and (\ref{eq:sigma_tilde_QS}).
}
\label{fig:simulation_results_1}
\end{figure}

We note that Eq.~(\ref{eq:P_T_QS_2}) is the energy-time filtering equation derived in Ref.~\cite{locane2019time} under the assumption that the time dependency of the barrier potential is linear. Our derivation is easily generalized to the case of nonlinear time dependence as long as the quasistatic condition of $\partial u/\partial t \ll \Delta_b^2/\hbar$ is satisfied at all time. Hence, our result shows that the quasistatic condition is another sufficient condition for the validity of the filtering form, Eq.~\eqref{eq:P_T_QS_2}. 

We finally obtain the R-ATD, by plugging Eq.~\eqref{eq:P_T_QS_2} into Eq.~\eqref{eq:P_tilde_ATD}, and using a change of variable $t_a \to t_a + (\epsilon - \bar{\epsilon})/\dot{u}$, $T_{\text{fro}}[\epsilon, t_a + (\epsilon - \bar{\epsilon}) / \dot{u}; t] = T_{\text{fro}}(\bar{\epsilon}, \bar{t}_a; \bar{t}_a+t - t_a)$, and $\bar{t}_a = -\bar{x}_0 / v_g$ is the mean arrival time at $x=0$ in the absence of the barrier,
\begin{equation}
  \tilde{P}_{\text{ATD}}(t)
  = \tilde{P}_{\text{ATD},\Delta_b \to 0}(t)
  * \frac{\partial T_{\text{fro}}(\bar{\epsilon}, \bar{t}_a;\bar{t}_a+ t)}{\partial t},
   \label{eq:P_tilde_ATD_QS_2}
\end{equation}
where $*$ denotes the convolution in time $t$, $f(t) * g(t) = \smallint^{\infty}_{-\infty} dt' \, f(t') g(t-t')$. $\tilde{P}_{\text{ATD}, \Delta_b \to 0}(t)$ is the R-ATD if the detector barrier had vanishing energy broadening. 
Therefore, this distribution is given by Eqs.~(\ref{eq:P_tilde_ATD_SC}) and (\ref{eq:sigma_tilde_SC}) and determined semiclassically, i.e., in a way that the quantum effect is only considered in the initial Wigner function, as discussed in Sec.~\ref{sec:SC}.
The term $\partial_t T_{\text{fro}}(\bar{\epsilon}, \bar{t}_a; \bar{t}_a +t)$ is the sensitivity of the transmission probability with respect to the time delay evaluated at the mean energy and arrival time of the packet;
it is a normalized distribution in time $t$ in a form of peak of width $\Delta_b/\dot{u}$ located around $t=0$.
Using the property of the convolutional form and Eq.~(\ref{eq:sigma_tilde_SC}),
the standard deviation of the R-ATD satisfies
\begin{equation}
\tilde{\sigma}_t = \sqrt{ \sigma_t^2 + (\sigma_\epsilon/\dot{u})^2 + (\Delta_b/\dot{u})^2} .
\label{eq:sigma_tilde_QS}
\end{equation}
Therefore, the R-ATD reproduces the ATD well when $\text{max}(\Delta_b, \sigma_\epsilon)/\dot{u} \ll \sigma_t$.

The numerical simulations confirm the quasistatic approximations.
Figure~\ref{fig:simulation_results_1}(a) shows the R-ATD $\tilde{P}_{\text{ATD}}(t)$ for various speeds $\dot{u}$ of the barrier.
The result of the numerical simulation (solid lines) is in a good agreement with the approximation~\eqref{eq:P_tilde_ATD_QS_2} (circle).
The R-ATD $\tilde{P}_{\text{ATD}}(t)$ approach the ATD (dashed) as the barrier speed becomes larger than $\Delta_b/\sigma_t$, while satisfying Eq.~(\ref{eq:sigma_tilde_QS}) as shown in Fig.~\ref{fig:simulation_results_1}(b).

Note that in the regime both satisfying the quasistatic and semiclassical conditions [see Fig.~\ref{fig:setup}(d)], the results~(\ref{eq:P_tilde_ATD_QS_2}) and (\ref{eq:sigma_tilde_QS}) are equivalent to Eqs.~(\ref{eq:P_tilde_ATD_SC}) and (\ref{eq:sigma_tilde_SC}), respectively.

\section{Conclusion}\label{sec:Conclusion}

We studied the tunneling of the electron through the dynamical potential barrier which operates to reconstruct the ATD.
We focus on quasistatic and semiclassical regimes and obtain analytical relation between the ATD and the R-ATD, as well as the relation between the standard deviations between the two distributions.
The results suggest that the condition for a successful reconstruction of the ATD is that the times for the barrier to rise its height by amounts of tunneling energy window of the barrier and packet energy uncertainty should be smaller than time uncertainty of the packet, namely $\text{max}(\Delta_b, \sigma_\epsilon)/\dot{u}\ll \sigma_t$.

In experimental situations, a successful reconstruction of the ATD can be achieved either in semiclassical or quasistatic regimes. 
The semiclassical regime can be achieved using a wide barrier so that $\Delta_b \ll \sigma_\epsilon $.
The successful reconstruction of the ATD requires a barrier speed to be fast as
$\dot{u} \gg \sigma_\epsilon/\sigma_t $.
The lower bound is $\sim$1 meV/ps for the Gaussian wave packet of $\sigma_\epsilon\sim 0.6$ meV~\cite{ryu2016ultrafast}, which is realistic~\cite{fletcher2019continuous}.
In the quasistatic regime,
the issue is whether a barrier slow enough for the quasistatic condition can be fast enough for the reconstruction of the ATD,
i.e., whether one can achieve $ \text{max}(\Delta_b, \sigma_\epsilon)/ \sigma_t  \ll \dot{u} \ll \Delta_b^2 / \hbar $.
This condition can be achieved, e.g., $\text{max}(\Delta_b, \sigma_\epsilon)/ \sigma_t \sim 10$ meV/ps, $\Delta_b^2 / \hbar \sim 60$ meV/ps for $\Delta_b \sim 6$ meV~\cite{ubbelohde2015partitioning}, and $\sigma_{\epsilon} = 0.6 \text{ meV}$~\cite{ryu2016ultrafast}.

We comment that the semiclassical regime $\Delta_b \ll \sigma_\epsilon$ may offer a merit over the opposite regime $\Delta_b \gg \sigma_\epsilon$, in a practical view in experimental situations.
Because, for given wave packet, the required barrier speed $ \text{max}(\Delta_b,\sigma_\epsilon)/\sigma_t$ is small for the former regime. 
Moreover, in the semiclassical regime, the same barrier can be used as an energy detector when operated statically with small measurement error $ \sim O( \Delta_b /\sigma_\epsilon)$~\cite{ryu2016ultrafast}.
Therefore, this regime is optimal for measuring the minimum Heisenberg uncertainty product for Gaussian wave packet in experiments~\cite{ryu2016ultrafast}.

Our work complements previous works of Ref.~\cite{locane2019time}, focusing on the measurement of the ATD. Our work suggests the analytic relation between the R-ATD with the ATD, and quantifies the error, while Ref.~\cite{locane2019time} suggests the relation between the packet transmission probability and a modified Wigner distribution which only equals the Wigner distribution when the height of the detector barrier increases linearly in time.

  Note that the traversal time does not play a role on the error of ATD reconstruction in the quasistatic or semiclassical regime;
  in the former regime, the effect of the photoassisted tunneling is weak, and in the latter regime, it lacks the quantum tunneling effect.

 In the regime which is neither quasistatic nor semiclassical, the traversal time can generally affect the time evolution of the wave packet during the tunneling.
  Unexpectedly,
  our numerical results suggest that the R-ATD is still well described by the results~(\ref{eq:P_tilde_ATD_QS_2}) and (\ref{eq:sigma_tilde_QS}), obtained in the quasistatic regime.
  This is demonstrated by the good agreement between the numerical result and quasistatic approximation in Fig.~\ref{fig:simulation_results_1}(b) for the parameters of $\dot{u} \sigma_t /\Delta_b > 5$. We attribute these behaviors to the fact that the effect of the photoassisted tunneling for the packet is negligible because the maximum tunneling time $\hbar/\Delta_b$ is much smaller than the temporal uncertainty of the packet $\sigma_t$.

Hence, the deviation from the quasistatic approximation occurs only in restricted parameter space in which $\hbar/\Delta_b \sim \sigma_t$; note that when $\hbar/\Delta_b \ll \sigma_t$ the photoassisted tunneling effect does not affect the packet much (we recall that $\hbar/\Delta_b$ is the maximum traversal time), and when $\hbar/\Delta_b \gg \sigma_t$ the quantum tunneling effect is weak as being in the semiclassical regime.
Indeed, we observe that the packet tunneling behavior can deviate from the quasistatic approximation in the restricted parameter space (see Appendix~\ref{sec:Breakdown_QS}).
However, the degree of the deviation from the quasistatic approximation is not large.
With these observations, we conclude that a condition for the successful reconstruction of the ATD is $\max{(\Delta_b, \sigma_\epsilon)/\dot{u}} \ll \sigma_t$, generally.

\section*{Acknowledgement}
This work is supported by Korea NRF (SRC Center for Quantum Coherence in Condensed Matter, Grants No. RS-2023-00207732 and No. 2023R1A2C2003430).
S.R. acknowledges a partial support from Universitat de les Illes Balears, FEDER, and NextGeneration EU.

\appendix

\section{Effective 1D Hamiltonian}
\label{sec:eff_1D}

Here we present the details about how the effective one-dimensional Hamiltonian of Eq.~\eqref{eq:Hamiltonian} describes the two-dimensional electron system (2DES) under a strong perpendicular magnetic field, following the approach of Sec.~III D in Ref.~\cite{huckestein_scaling_1995}.

We consider a 2DES under a strong perpendicular magnetic field of $\vec{B} = \pm |B| \hat{z}$ such that the electrostatic potential $V_{\text{2D}}$ changes smoothly over the magnetic length $l_B $ ($=\sqrt{\hbar/e|B|} $) and slowly over the cyclotron frequency $\omega_c$.
Then, the Landau level index $n$ becomes a constant of motion and the Hamiltonian of 2DES is written as~\cite{huckestein_scaling_1995}
\begin{equation}
  H_{\text{2D}} = (n + \frac{1}{2}) \hbar\omega_c + V_{\text{2D}}(X, Y; t) ,
\end{equation}
where $X$ and $Y$ are the guiding-center position operators, satisfying the canonical commutation relation
\begin{equation}
  [X, Y] = \pm i l_B^2 .
\end{equation}
Here the upper signs in the $\pm$ and $\mp$ factors shown in this Appendix are for the magnetic field $\vec{B}=|B| \hat{z}$, while the lower signs are for the opposite direction $\vec{B}=- |B| \hat{z}$.

We describe the electrostatic potential as the sum of the harmonic edge confinement $(\sim m_e^* \omega_y^2 Y^2/2 )$ and the potential barrier which is inverse harmonic with a barrier smoothing by the gate,
\begin{equation}
  V_{\text{2D}} (X,Y; t)= \frac{1}{2}m_e^* \omega_y^2 Y^2
  + f_{s}(X) \Big(u(t;\mathbb{T}) - \frac{1}{2} m_e^*\omega_x^2 X^2\Big) .
\end{equation}
$m_e^*$ is the 2DES effective electron mass.
$f_{s}(X)$,  $u(t;\mathbb{T})$, and $\mathbb{T}$ are the barrier smoothing factor, the height of the barrier, and the time delay, respectively as introduced in Sec.~\ref{sec:potential}.
Note that a previous experiment~\cite{kataoka2016time} has shown that the harmonic edge confinement well describes the quadratic dispersion relation between the energy and momentum of the hot electrons (whose energy is typically $\sim$100 meV above the Fermi energy) propagating along the edge.
In the linear response regime, where the charge transport is determined by the electrons of energy near the Fermi energy (typically within 0.1 meV for a temperature below 1 K), the dispersion relation effectively reduces to the linear one.

Due to the canonical commutation relation between $X$ and $Y$, the Hamiltonian can be written only in terms of $X$ by substituting $Y \rightarrow \mp i l_B^2 \partial_X$,
\begin{align}
    H_{\text{2D}}
  &= (n+\frac{1}{2})\hbar\omega_c -\frac{\hbar^2}{2m^*}\frac{\partial^2}{\partial X^2}+U(X,t;\mathbb{T})  \label{eq:H2D-X}
  \\
  U(X,t;\mathbb{T})
    &= f_{s}(X) \Big(u(t;\mathbb{T}) - \frac{1}{2} m^*\omega_b^2 X^2\Big) , 
\end{align}
where $m^*= m_e^* \omega_c^2/\omega_y^2$, $\omega_b = \omega_x\omega_y/\omega_c$, and $\omega_c = e|B|/m_e^*$. 
Equation (\ref{eq:H2D-X}) is the one-dimensional Hamiltonian which describes the time evolution of the guiding center by the Heisenberg equation of motion (when interpreting $\pm \hbar Y/l_B^2$ as the momentum)
\begin{align}
  \dot{X} &= \pm \dfrac{\hbar Y}{m^* l_B^2} = \pm \dfrac{l_B^2}{\hbar} \dfrac{\partial U}{\partial Y}, \\
  \dot{Y} &= \mp \dfrac{l_B^2}{\hbar} \dfrac{\partial U}{\partial X}.
\end{align}
These equations describe the $E$-cross-$B$ drift motion. When measuring the energy from the bottom of Landau level subband, Eq.~(\ref{eq:H2D-X}) corresponds to Eq.~(\ref{eq:Hamiltonian}).

\section{Correlated energy-time distribution}
\label{sec:Correlated_state}

\begin{figure}[t]
\centering
\includegraphics[width=.5\linewidth]{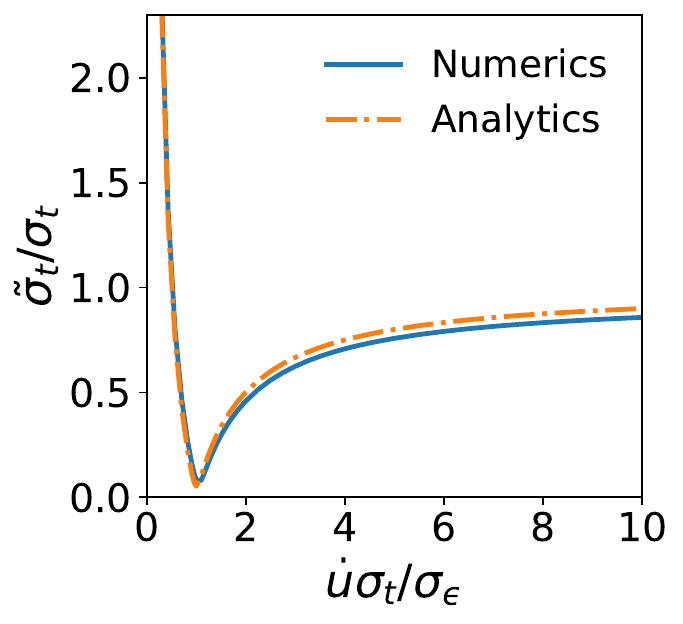}
\caption{The time uncertainty $\tilde{\sigma}_t$ of the R-ATD as a function of the barrier speed $\dot{u}$ in the case of the correlated energy-time distribution, Eq.~\eqref{eq:Correlated_state}. 
  The analytic result (dashed-dotted curve), Eq.~\eqref{eq:sigma_tilde_SC_2}, is compared with the numerical simulation (solid) in the semiclassical regime.
  Parameters:  $\sigma_t = 2.5 \text{ ps}$ and $\alpha = 1 \text{ meV$/$ps}$, which gives $\sigma_\epsilon \sim 2.5 \text{ meV}$ in consistency with Ref.~\cite{kataoka2017time}. $\Delta_b = 0.1 \text{ meV}$, $\bar{t}_a = 10 \text{ ps}$, and $\bar{\epsilon} = 100 \text{ meV}$.
}
\label{fig:simulation_results_3}
\end{figure}

Here we discuss how the relation~(\ref{eq:sigma_tilde_SC}) is modified for non-Gaussian wave packet. Let us consider a packet correlated in the energy-time space which can be generated in the quantum-dot pumps with slow emission protocol~\cite{kataoka2017time, fletcher2019continuous}.
The pure-state initial Wigner distribution can be described as
\begin{equation}
\mathcal{W}_0(t_a ,\epsilon) = \dfrac{1}{\pi \hbar} \exp \bigg[ - \dfrac{(t_a - \bar{t}_a)^2}{2 \sigma_t^2} - \dfrac{[\epsilon - \bar{\epsilon} - \alpha (t_a-\bar{t}_a)]^2}{2 [\hbar / (2 \sigma_t)]^2 } \bigg],
\label{eq:Correlated_state}
\end{equation}
where $\alpha$ describes the correlation between the time and energy, $\sigma_t$ is the temporal uncertainty, and the energy uncertainty follows:
\begin{equation}
  \sigma_{\epsilon} = \sqrt{ [\hbar /(2\sigma_t)]^2 + (\alpha \sigma_t)^2 }.
  \label{eq:sigE-corr}
\end{equation}
We choose positive $\alpha$ as in the experimental situation~\cite{kataoka2017time}.

We calculate an analytical form of the standard deviation $\tilde{\sigma_t}$ of the R-ATD in the semiclassical regime, $\Delta_b \ll \sigma_\epsilon$. Using Eqs.~\eqref{eq:P_tilde_ATD_SC} and \eqref{eq:Correlated_state}, we find that
\begin{equation}
  \tilde{\sigma}_t
  = \sqrt{ \sigma_t^2 ( 1 - 2 \alpha/\dot{u} ) + (\sigma_\epsilon / \dot{u})^2 } .
\label{eq:sigma_tilde_SC_2}
\end{equation}
$\tilde{\sigma_t}$ approaches the standard deviation $\sigma_t$ of ATD when $\dot{u} \gg \sigma_{\epsilon}/\sigma_t$ because this condition also guarantees $\dot{u} \gg \alpha$ due to Eq.~(\ref{eq:sigE-corr}).

We confirm Eq.~(\ref{eq:sigma_tilde_SC_2}) by direct numerical computation (see Appendix~\ref{sec:Numerical_simulations} for the method) of the time evolution of the wave packet (see Fig.~\ref{fig:simulation_results_3}).
Note that $\tilde{\sigma}_t$ can be smaller than $\sigma_t$ [with minimum occurring at $\dot{u} = \sigma_{\epsilon}^2 /(\alpha \sigma_t^2)$] due to the correlation in the energy-time space, as demonstrated in experiments~\cite{kataoka2017time}.

\section{Barrier traversal time}
\label{sec:Traversal_time}

\begin{figure}[t]
\centering
\includegraphics[width=.95\columnwidth]{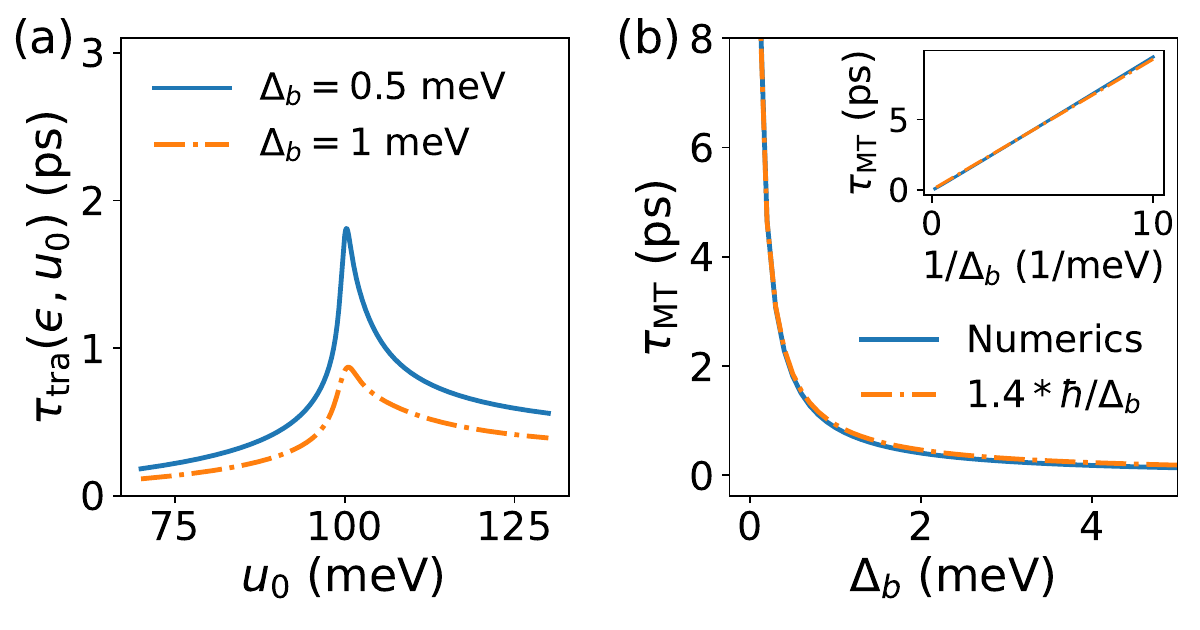}
\caption{Traversal time of the static potential barrier of the inverse harmonic shape with $u(t;\mathbb{T}) = u_0$ in Eq.~\eqref{eq:potential_inverse_harmonic}. The incident plane wave has the energy of $\epsilon=100 \text{ meV}$.
  (a) The traversal time as a function of the barrier height $u_0$. Two different values of $\Delta_b=$ 0.5 meV (solid line) and 1 meV (dashed-dotted line) are chosen. The traversal time is maximal around $u_0 = \epsilon$.
  (b) The maximum traversal time $\tau_{\text{MT}}$ over $u_0$ is drawn as a function of $\Delta_b$.  The numerical simulation (solid line) is in good agreement with $1.4\hbar/\Delta_b$ (dashed-dotted line). Note that $l_{s} = 10 l_b$ is chosen; the choice of $l_{s}$ does not alter the maximum traversal time since the transversal time is determined by electron motion near the top of the barrier.  
  Inset: $\tau_{\text{MT}}$ as a function of $1/\Delta_b$. }
\label{fig:traversal_time}
\end{figure}

We consider a static barrier of the inverse harmonic shape with barrier height $u(t;\mathbb{T}) = u_0$  in Eq.~\eqref{eq:potential_inverse_harmonic}, and show that the traversal time [see Eq.~(\ref{eq:traversal_time})] for transmission of an electron in a plane wave through the barrier is maximal when the energy $\epsilon$ of the electron is almost aligned with the barrier height and that the maximal value is $\sim \hbar/\Delta_b$. 
For this purpose, we calculate the transmission amplitude and the traversal time by obtaining a single-particle Green function numerically~\cite{datta1997electronic}.
 
Figure~\ref{fig:traversal_time}(a) shows the traversal time $\tau_{\text{tra}}(\epsilon, u_0)$ as a function of the barrier height $u_0$.
The traversal time becomes maximal at $u_0 \simeq \epsilon$.
The behavior can be understood as follows.
When $u_0< \epsilon$,  the traversal time decreases as $u_0$ decreases since the electron velocity increases near the barrier.
When $u_0> \epsilon$, the traversal time decreases as $u_0$ increases because the imaginary momentum in the tunneling increases~\cite{buttiker1982traversal}.
Figure~\ref{fig:traversal_time}(b) shows that the maximum traversal time follows $\hbar/\Delta_b$ up to a factor of order $1$.

\section{Validity of the semiclassical approximation}
\label{sec:QC}

We show the validity of the semiclassical approximation used in Sec.~\ref{sec:SC}, by showing that the second-order quantum correction in the semiclassical approximation is negligible.

The semiclassical approximation of the packet transmission probability, Eq.~(\ref{eq:P_T_SC_0}), is the leading-order term in the path-integration formulation and the expansion in the powers of $\hbar$. The next-order correction $\delta P_T$ to the packet transmission probability $P_T$ in Eq.~(\ref{eq:P_T_SC_0}) is determined by a nonharmonicity of the potential $U$ as~\cite{polkovnikov2010phase}
\begin{align}
  \begin{split}
    \delta P_{T}  =
    & \int_0^\infty d\tau \int^{\infty}_{-\infty} dx \int^{\infty}_{-\infty} dp \,
      W^{\text{(sc)}}_\tau(x, p)  \\
    & \qquad \qquad \times  \dfrac{\hbar^2}{24}
      \dfrac{\partial^3 U(x, \tau) }{\partial x^3}
      \dfrac{\partial^3 P_{T,\text{cl}}(x,p,\tau) }{\partial p^3}  .
      \label{eq:P_SC1}
  \end{split}
\end{align}
$W_\tau^{\textrm{(sc)}} (x,p) $ is the Wigner distribution at position $x$, momentum $p$, and time $\tau$, which follows the classical Liouville equation. $P_{T,\text{cl}}(x, p, \tau) $ describes whether a particle classically evolved from the position $x$, momentum $p$, and time $\tau$ ultimately passes over the barrier after a long-time evolution or not; it is 1 when the particle passes and 0 otherwise. 
The second line of Eq.~(\ref{eq:P_SC1}) describes the change of the transmission probability due to the quantum fluctuation induced by the nonharmonicity of the potential at position $x$, momentum $p$, and time $\tau$.
$P_{T,\text{cl}}$ can be written as $\Theta[p - p^*(x, \tau)]$, where $p^*(x,\tau)$ is the threshold value for the momentum above which the particle at position $x$ and time $\tau$ passes the barrier. Using the derivative of the Heaviside step function $\Theta[\cdots]$, and the integration by parts, Eq.~(\ref{eq:P_SC1}) is written as 
\begin{align}
\begin{split}
\delta P_{T}  = \int^{\infty}_0 d \tau &  \int^{\infty}_{-\infty} dx \, \dfrac{\hbar^2}{24} \dfrac{\partial^3 U(x, \tau) }{\partial x^3} \\ & \times \dfrac{\partial^2}{\partial p^2} W_\tau^{\textrm{(sc)}} (x, p)\Big|_{p = p^*(x, \tau)} .
\label{eq:P_SC2}
\end{split}
\end{align}

Now we estimate the order of Eq.~(\ref{eq:P_SC2}).
The time (respectively position) interval for nonvanishing integrand is $O(l_{s}/v_g)$ [$O(l_{s})$] because the nonharmonicity of the potential only exists over the length scale $l_{s}$.
The third-order derivative of the potential is determined by the barrier parameters as $ \partial^3 U(x, \tau) /\partial x^3 = O (\Delta_b/l_{s}^3) $.
The second-order derivative of the Wigner function is determined by the initial energy uncertainty $\sigma_\epsilon$ as
$\partial^2 W^{(\text{sc})}_\tau/\partial p^2 = O [v_g^2/(\hbar  \sigma_\epsilon^2)] $.
Multiplying all the factors, we obtain
\begin{equation}
  \delta P_T
  = O \Big( \frac{\Delta_b}{\sigma_\epsilon}
    \frac{\hbar v_g}{l_{s} \sigma_\epsilon}\Big) .
\end{equation}
 $\Delta_b/\sigma_\epsilon$ describes the portion of the wave packet which undergoes probabilistic quantum tunneling through the barrier, as discussed in Sec.~\ref{sec:Regimes}. $\hbar v_g / \sigma_{\epsilon}$ is the length scale of the wave packet and $l_s$ is the length scale of the nonharmonic potential. Hence, $\hbar v_g/(l_{s} \sigma_\epsilon)$ describes the strength of quantum fluctuation which may arise due to nonharmonicity of the potential. 
This term is typically in the order of 1 in the experimental situations, e.g.,  
$v_g = 10^5 \text{ m/s}$~\cite{kataoka2016time}, $\sigma_{\epsilon} = 0.6 \text{ meV}$~\cite{waldie2015measurement},  $l_{s} \sim 100 \text{ nm}$ (roughly estimated as the depth of 2DES), hence $\hbar v_g/(l_{s} \sigma_\epsilon)\sim 1$.
Therefore, we confirm that the condition $\Delta_b \ll \sigma_\epsilon$ indeed guarantees the validity of the semiclassical approximation.

\begin{figure}[t!]
\centering
\includegraphics[width=.95\columnwidth]{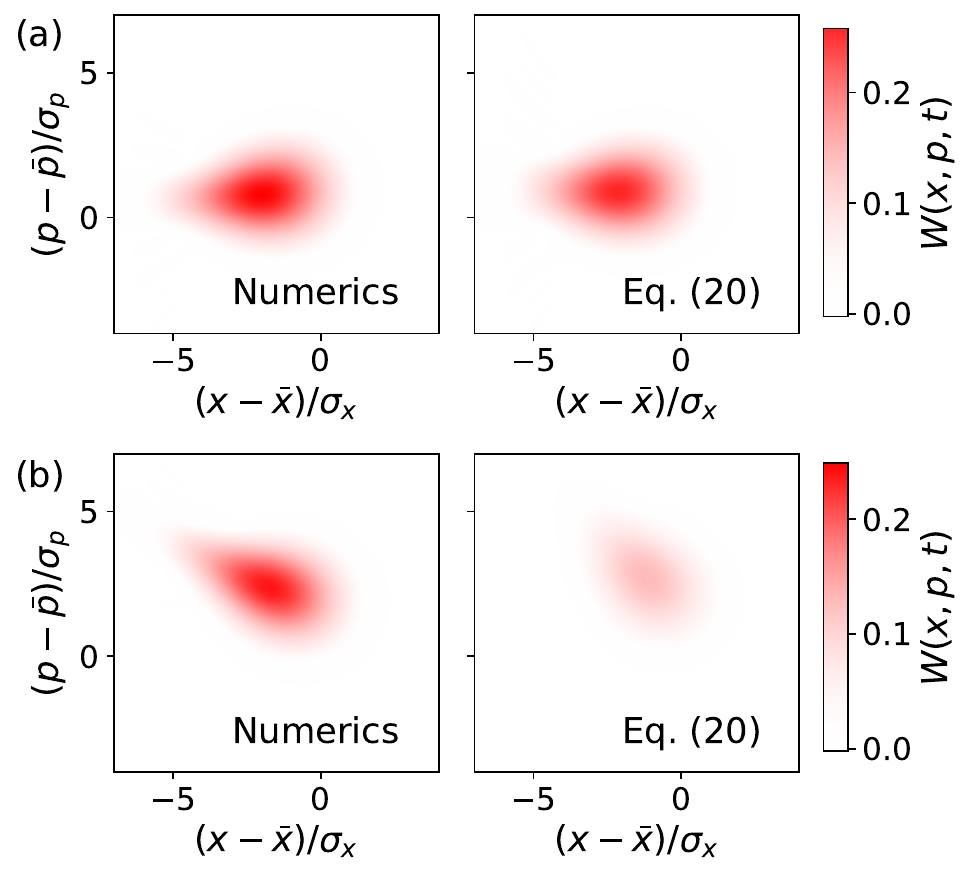}
\caption{Wigner distributions $W(x,p)$ of the transmitted electron in the quasistatic regime (a) and the nonquasistatic regime (b).
  The results from the numerical simulation (left) are compared with the quasistatic approximation~\eqref{eq:psi_T_QS} (right).
  Parameters: $\dot{u} \hbar/ \Delta_b^2 = 0.18$ for (a) and 1.46 for (b).
For both (a) and (b), $\Delta_b = \sigma_\epsilon = 0.6 \text{ meV}$, $\mathbb{T} = -\bar{x}_0/v_g $, and the incident wave packet is Gaussian.}
\label{fig:wigner_distribution_2}
\end{figure}

\section{Numerical time evolution of packet}
\label{sec:Numerical_simulations}

Here we present the method for the numerical simulations used to obtain results of Figs.~\ref{fig:simulation_results_2}--\ref{fig:simulation_results_3} and \ref{fig:wigner_distribution_2}. To numerically solve the time-evolution of the wave function according to the time-dependent Hamiltonian~\eqref{eq:Hamiltonian}, we discretize the Hamiltonian in position and time~\cite{datta1997electronic}.

The lattice constant $a$ is chosen such that the energy band is well described by a quadratic relation $\epsilon(k) = \hbar^2 k^2/(2m^*)$. We choose $a = 0.2 \text{ nm}$ which satisfies the quadratic relation well within the energy of the packet. For the chosen value of $a$, the energy of the wave packet $\sim$100 meV is much smaller than the maximum energy of the discretized band $2 \hbar^2/(m^* a^2) \sim 1180 \text{ meV}$.

The time evolution of the wave function is then obtained by numerically solving the time-dependent Schr\"{o}dinger equation using fourth-order Adams-Bashforth algorithm~\cite{butcher2008numerical}.
For an accurate result, we choose the time step $\delta =10^{-4} $ ps so that it becomes much smaller than the fastest timescale of the discretized model, i.e.,  $2m^*a^2/(2\hbar) \sim 10^{-3}$ ps.
We simulate the time evolution until the moment that the probability that the electron on the right side of the barrier converges.

We choose $m^*\sim 50 m^*_e$ and $\bar{\epsilon}=100$ meV so that the group velocity $v_g = 10^5 \text{ m/s}$ becomes similar to experimentally reported value~\cite{kataoka2016time}.
The choice is also consistent with the parameters of the two-dimensional situation, e.g. $m^*=m_e^* \omega_c^2/\omega_y^2 \sim 50 m^*_e$ and $\hbar \omega_c\sim 20$ meV give $\hbar \omega_y \sim 3 $ meV in consistency with the experiment.
Note that the detailed values for the choices do not change our results about the relation between the ATD and the R-ATD.
The distance between the center of the initial wave packet and the top of the potential barrier is chosen as $ |\bar{x}_0| = 1 \text{ $\mu$m}$ so that the travel time to the barrier center $|\bar{x}_0|/v_g = 10 \text{ ps}$ is sufficiently larger than packet temporal width but sufficiently small to suppress the nonlinear dispersion effect.

\section{Breakdown of quasistatic approximation}
\label{sec:Breakdown_QS}

In this appendix, we show that the quasistatic approximation~\eqref{eq:psi_T_QS} breaks down when $\dot{u}> \Delta_b^2/\hbar$ and $\sigma_\epsilon = \Delta_b$.

Figure~\ref{fig:wigner_distribution_2} shows Wigner distributions of transmitted wave packets for the quasistatic regime and non-quasistatic regime. In the quasistatic regime [Fig.~\ref{fig:wigner_distribution_2}(a)], the numerical simulation shows a good agreement with the quasistatic approximation.
In the non-quasistatic regime [Fig.~\ref{fig:wigner_distribution_2}(b)], the numerical simulation differs from the quasistatic approximation. In the numerical simulation, we observe (i) larger transmission probability, i.e. $\smallint^{\infty}_0 dx \smallint^{\infty}_{-\infty} dp \, W(x,p,t)$, and (ii) longer tail of the packet, compared to those in the semiclassical approximation. These differences occur since the frozen transmission amplitude $d_{\text{fro}}$ in the semiclassical approximation does not describe the photoassisted tunneling; such tunneling can enhance the packet transmission in the nonquasistatic regime.


%

\end{document}